\documentclass[sigconf]{acmart}

\usepackage{subfig} 

\AtBeginDocument{%
  \providecommand\BibTeX{{%
    \normalfont B\kern-0.5em{\scshape i\kern-0.25em b}\kern-0.8em\TeX}}}

\setcopyright{acmcopyright}
\copyrightyear{2018}
\acmYear{2018}
\acmDOI{XXXXXXX.XXXXXXX}

\acmConference[Conference acronym 'XX]{Make sure to enter the correct
  conference title from your rights confirmation emai}{June 03--05,
  2018}{Woodstock, NY}
\acmPrice{15.00}
\acmISBN{978-1-4503-XXXX-X/18/06}

\begin{document}

\title{Mobile Supply: The Last Piece of Jigsaw of Recommender System}

\author{Zhenhao Jiang}
\authornote{Equal contribution.\\This paper was done when Zhenhao Jiang as an intern in Alibaba Group.}
\affiliation{%
  \institution{The Chinese University of Hongkong,}
  \institution{Shenzhen Research Institute of Big Data}
  \city{Shenzhen}
  \country{China}
}
\email{222041010@link.cuhk.edu.cn}

\author{Biao Zeng}
\authornotemark[1]
\affiliation{%
  \institution{Alibaba Group}
  \city{Hangzhou}
  \country{China}}
\email{biaozeng.zb@alibaba-inc.com}

\author{Hao Feng}
\affiliation{%
  \institution{Alibaba Group}
  \city{Shanghai}
  \country{China}}
\email{zhisu.fh@alibaba-inc.com}

\author{Jin Liu}
\authornote{Corresponding authors.}
\affiliation{%
\institution{Alibaba Group}
  \city{Hangzhou}
  \country{China}}
\email{nanjia.lj@alibaba-inc.com}

\author{Jie Zhang}
\affiliation{%
\institution{Alibaba Group}
  \city{Shanghai}
  \country{China}}
\email{shenxu.zj@alibaba-inc.com}

\author{Jia Jia}
\affiliation{%
\institution{Alibaba Group}
  \city{Shanghai}
  \country{China}}
\email{jj229618@alibaba-inc.com}

\author{Ning Hu}
\affiliation{%
\institution{Alibaba Group}
  \city{Shanghai}
  \country{China}}

\renewcommand{\shortauthors}{Jiang and Zeng, et al.}

\begin{abstract}
  Recommendation system is a fundamental functionality of online platforms. With the development of computing power of mobile phones, some researchers have deployed recommendation algorithms on users' mobile devices to address the problems of data transmission delay and pagination trigger mechanism. However, the existing edge-side mobile rankings cannot completely solve the problem of pagination trigger mechanism. The mobile ranking can only sort the items on the current page, and the fixed set of candidate items limits the performance of the mobile ranking. Besides, after the user has viewed the items of interest to the user on the current page, the user refresh to get a new page of items. This will affect the user's immersive experience because the user is not satisfied with the left items on the current page. In order to address the problem of pagination trigger mechanism, we propose a completely new module in the pipeline of recommender system named Mobile Supply. The pipeline of recommender system is extended to ``retrival->pre-ranking->ranking->re-ranking->Mobile Supply->mobile ranking". Specifically, we introduce the concept of list value and use point-wise paradigm to approximate list-wise estimation to calculate the maximum revenue that can be achieved by mobile ranking for the current page. We also design a new mobile ranking approach named device-aware mobile ranking considering the differences of mobile devices tailored to the new pipeline. Extensive offline and online experiments show the superiority of our proposed method and prove that Mobile Supply can further improve the performance of edge-side recommender system and user experience. Mobile Supply has been deployed on the homepage of a large-scale online food platform and has yielded considerable profits in our business.
\end{abstract}



\received{20 February 2007}
\received[revised]{12 March 2009}
\received[accepted]{5 June 2009}

\maketitle

\section{Introduction}
Shipping proper items to users based on users' preferences is the basic functionality of recommendation systems that are fundamental parts of online platforms like e-commerce, short-video, online food platform, \textit{etc.} \cite{yu2021dual,du2022basm}. Capturing the user's interest is the core task of recommendation systems, but the user's interest changes constantly. In order to extract user interest more accurately, the recommendation systems should respond quickly to the real-time behavior of the user. As a result, it is very important for the recommender systems to be more sensitive to user's real-time feedback. However, the traditional recommendation systems cannot do real-time recommendation because of data transmission delay and pagination trigger mechanism \cite{gong2020edgerec}. Briefly speaking, the traditional recommendation systems compute the recommendation list on the cloud-side and then send it to the user's device via the network. The user behavior data also needs to be uploaded to the cloud-side via the network. This data transmission delay makes real-time recommendations impossible. In addition, the traditional recommendation systems employ the pagination trigger mechanism, only when the user performs refresh or page turning operation, the new recommendation list is sent to the user's device, which makes it impossible to adjust the recommendation list in real-time according to the user's behaviors (\textit{e.g.} click, favorite).

As the computing power of mobile devices increases, it becomes possible to store and use deep learning models on mobile devices \cite{zhang2018shufflenet, pan2022edgevits}. When the ranking model is deployed on the user's mobile device, the model‘s inference no longer has the data transmission delay. Besides, instead of waiting for page turning operation, the recommendation list can be adjusted in real-time based on user behavior by the mobile ranking. Therefore, the mobile ranking can do real-time recommendation. However, there is a problem with the existing mobile rankings.

\begin{figure*}[htbp]
	\centering
	\subfloat[]{\includegraphics[width=.26\linewidth]{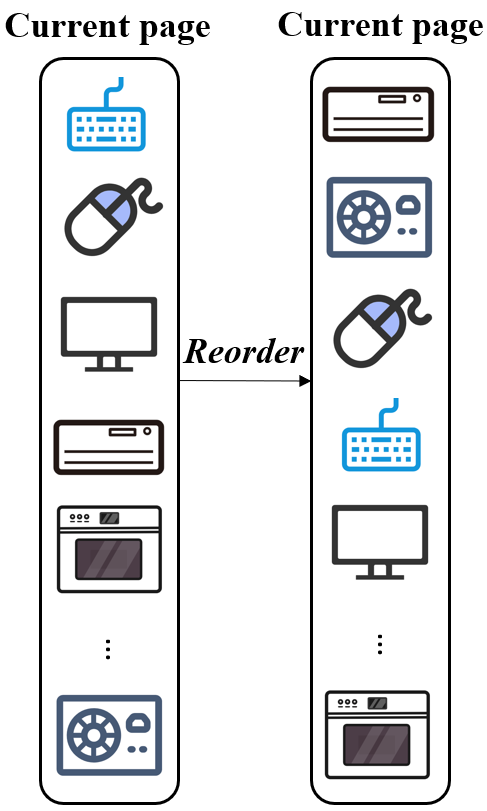}\label{fig1b}}\hspace{6pt}
	\subfloat[]{\includegraphics[width=.33\linewidth]{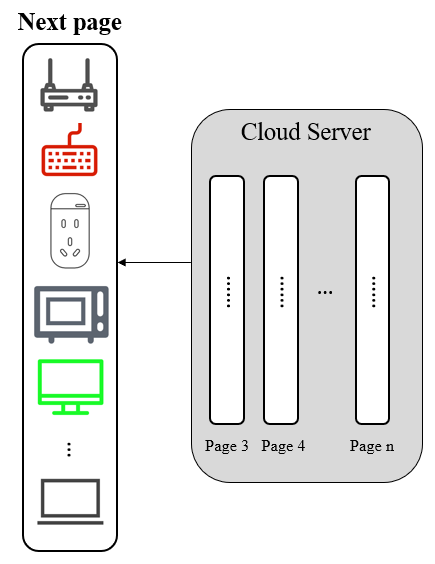}\label{fig1c}}\hspace{6pt}
	\subfloat[]{\includegraphics[width=.36\linewidth]{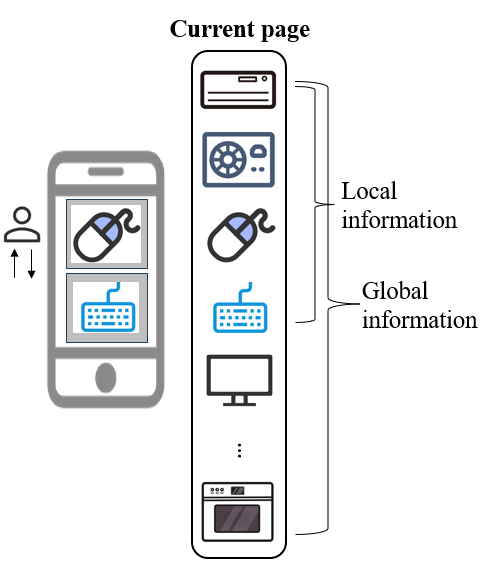}\label{fig1a}}
	\caption{Illustration of mobile recommender. (a) Mobile ranking reorders the items on the current page. But the items contained in the current page are fixed. As a result, the performance of mobile ranking is limited. (b) Mobile Supply can automatically apply for the next page items from the server, these items are new to the user. Mobile Supply+mobile ranking can give users an immersive experience. (c) Item lists where local and global information resides. Users can swipe up and down to see what the current page contains on their phone.}
\end{figure*}

\textbf{Restrictions on candidate items.} Actually, there are two problems for the pagination trigger mechanism: 1) the recommendation list cannot be updated in real-time and 2) candidate items cannot be updated in real-time. The existing researches only discussed problem 1 but ignored problem 2. In reality, the entire recommendation lists are sent page by page from the cloud-side server to the user's device. This means that if the user does not trigger pagination, the cloud server will not send a new list, and the set of candidate items for mobile ranking on the user's device is fixed. Therefore, the mobile ranking can only adjust the order on a limited set of items, which severely limits the performance of mobile ranking shown in Fig.~\ref{fig1b}. Moreover, There are two spin-offs to this problem.

 \textbf{1) Affect the immersive experience.} Every time a user logs onto an online platform, the behavior on the platform can be regarded as a process of convergence of interest. However, the user's interest may converge beyond the scope of the current page, and when the current page fails to meet the user's interest, the user sends a paging request to get the next page. Therefore, the user has been disappointed with the current page when they send the paging request, affecting the user experience. \textbf{2) Limit browsing depth.} Because each page turning affects the user's interest, the user's stay time and browsing depth will decrease.
The decline in browsing depth leads to many items of interest to users being missed, so that users' willingness to stay in online platform is further reduced, forming a vicious circle.

To bridge this research gap, in this paper, we propose a new module in recommendation system named Mobile Supply. Here, the traditional pipeline of recommendation system is extended to ``retrival->pre-ranking->ranking->re-ranking->\textbf{Mobile Supply}->mobile ranking". Mobile Supply is deployed on the user's mobile device to determine whether to request the next page according to the user's real-time behavior. This is equivalent to an automatic pagination trigger. Mobile Supply can automatically send paging requests according to user's real-time behavior, and expand the collection of candidate items on the mobile device to better meet the convergence process of user interest shown in Fig.~\ref{fig1c}. In addition, paging can occur when the user is browsing an item in the in-shop page (item detail page), and the user will not perceive the page turning, allowing the user to have an immersive experience. Besides, Mobile Supply can be combined with mobile ranking to always expose items that the user is interested in, so that the user can maintain the browsing willingness and increase the browsing depth. 

Specifically, we consider the local information and the global information of the list of candidate items shown in Fig.~\ref{fig1a}. Due to the limited computing power of mobile devices, it is difficult to use list-wise methods to generate recommendation list. We directly employ a point-wise approach to estimate the sum of orders of recommendation list. The Mobile Supply mainly consists of three parts: 1) real-time interest extractor (RIE), 2) contextual information extractor (CIE), and 3) Uplift MMoE (UM). First, RIE extracts the user's real-time interest from the real-time behavior and real-time features, and CIE extracts contextual information from the list of candidate items in the current page. Once real-time interest vector and contextual information vector are obtained, they are concatenated to feed into UM. UM estimates the total values of the exposed list and candidate items list. Then, judge if sending the paging request is necessary based on their difference in value (uplift).

Moreover, we also give an introduction of our device-aware mobile ranking (DMR) based on scene MMoE \cite{he2020dadnn} to consider Android and IOS users separately. Because Mobile Supply is a completely new module in recommender systems, we also articulate the system architecture and emphasize the experience learned along the way to successfully deploy such system in a billion-user scale online food platform.

our contributions are summarized as follows:
\begin{itemize}
    \item We first propose a completely new module in recommender systems \textit{i.e.} Mobile Supply. The traditional pipeline of recommendation system is extended to ``retrival->pre-ranking->ranking->re-ranking->Mobile Supply->mobile ranking". The traditional pagination trigger mechanism is comprehensively upgraded to tailor real-time recommendation.
    \item We first propose a mobile ranking considering Android and IOS users separately.
    \item We share unique and valuable experience (system architecture, real-time feature system, model design, \textit{etc.}) of
    successful deploying Mobile Supply and mobile ranking on a large-scale online food platform serving billions of users.
    \item Extensive offline and online experiments on two real-world datasets demonstrate that Mobile Supply is critical for the pipeline of recommendation system. 
\end{itemize}

\section{related works}
\subsection{Point-Wise, Pair-Wise and List-Wise}
According to the different forms of samples, algorithms in recommendation are mainly divided into point-wise, pair-wise, and list-wise. 

\textbf{Point-wise} aims to predict whether a user will click or purchase an item (\textit{e.g.} Click-Through Rate (CTR)/Conversion Rate (CVR) estimation \cite{ma2018entire}) or predict rating of an item given by a user (\textit{i.e.} Movie Rating prediction \cite{li2020research}). It considers each item independently, focusing only on the interaction between the current item and the user. Because of the low-cost of computation, complex ranking models generally choose this way \cite{covington2016deep}. But in reality, the user does not consider different items independently, the user will make decisions based on all the items that can be seen on the screen.

\textbf{Pair-wise} aims to learn the semantic distance of a pair of items, whose core idea is to increase the sensitivity of the model to positive and negative samples \cite{koppel2020pairwise}. However, the recommendation list can only be locally optimal in this manner. This method is commonly used in retrival algorithms \cite{nigam2019semantic}.

\textbf{List-wise} aims to optimize the entire recommendation list directly from a global perspective \cite{he2017category}. Most list-wise ranking algorithms are generative, generating the entire recommendation list directly based on the characteristics of the collection of candidate items and other contextual information \cite{liu2023generative}. This method is the most computationally complex, but conforms to reality that is widely used in re-ranking \cite{feng2021grn}. 

Different from the above studies, our method uses a point-wise approach to approximate the performance of list-wise estimation. Not only the contextual information is considered, but also the complexity of the algorithm is reduced.

\subsection{Recommendation on Mobile Devices}
Traditional mobile recommender only includes mobile ranking which can be considered as a special re-ranking. It re-ranks the recommendation list given by the server based on the real-time behavior of user \cite{yao2021device}. With the development of computing power of mobile devices in recent years, practitioners have gradually explored the mobile ranking. At present, there are few papers on mobile recommendation. 

It is generally believed that EdgeRec \cite{gong2020edgerec} is the first mobile ranking algorithm to be deployed on a large-scale recommendation system. This method has achieved great success on Taobao (e-commerce platform). This algorithm uses GRU \cite{dey2017gate} to extract contextual information about candidate items and attention mechanism \cite{vaswani2017attention} to extract user interest. Going further, \cite{gu2022device} aims to train model with data of user and similar users retrieved from the server on user's device. 

KuaiRec \cite{gong2022real} is the latest mobile ranking with practical applications that is deployed on Kuaishou (short-video platform). This approach employs multi-head attention (MHA) \cite{vaswani2017attention} to extract user interest and MMoE \cite{ma2018modeling} to predict multiple targets simultaneously. In addition, it uses beam search \cite{freitag2017beam} to determine the impact of different orders on contextual information. There are also studies \cite{huang2022geographical,wang2020trusted} that consider combining Federated Learning \cite{zhang2021survey} and mobile ranking to protect users' privacy.

Different from the above studies, we design a completely new module named Mobile Supply aiming to further improve the performance of mobile ranking. In addition, we also introduce our device-aware mobile ranking model.

\section{system architecture}
\begin{figure}[t]
    \centering
    \includegraphics[width=0.95\columnwidth]{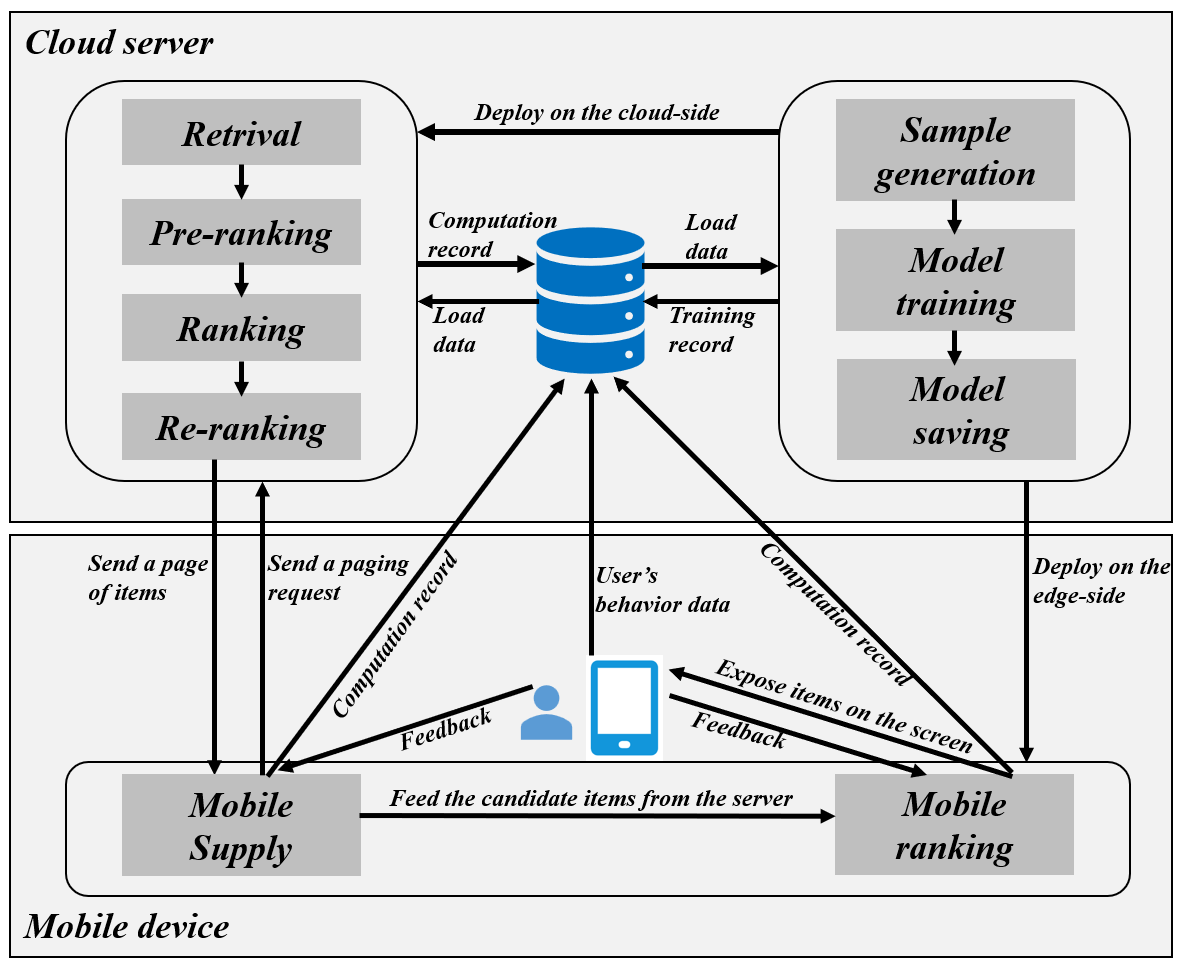}
    \caption{The pipeline of the large-scale industrial recommendation system.}
    \label{fig2}
    \Description{The relationship between server-side recommender and edge-side recommender and the information transfer path in cloud-edge collaborative recommendation system.}
\end{figure}

In this section, we introduce the pipeline of the large-scale industrial recommendation system. The whole framework can be divided into three parts shown in Fig.~\ref{fig2}.

\subsection{Cloud-Side Recommendation}
The traditional recommendation system is deployed on the cloud-side server which mainly consisting of retrival, pre-ranking, ranking and re-ranking. Through these stages, items are ranked and stored on the server. Once the server receive a pagination request, it will send one page of items to the user's device. There is a database in cloud-side that can store models' computation records and user's feedback to generate samples to train models. Mobile Supply and mobile ranking are also trained on the server. After training, mobile models will be deployed on the user's device with the update of application. Models on the server can only compute recommendation list with delayed user actions uploaded to the server.

\subsection{Edge-Side Recommendation}
The traditional edge-side recommendation system only includes mobile ranking. In our system, Mobile Supply is a member of edge-side recommendation system as well. Mobile Supply will calculate the uplift between exposed items and all candidate items to determine whether the candidate items in the current page meet the user's interest. If not, it sends a paging request to the server to obtain a new page of candidate items. In addition, users can still turn pages manually. After confirming the candidate items, mobile ranking re-ranks the items to expose them on the user's device. The computation records of mobile models are uploaded to the server.

\subsection{Real-Time Feature System}
Due to the limitation of transmission speed and mobile device storage capacity, we need to design an edge-cloud collaborative feature system. 

Real-time features come from three parts: item, server and device. The basic characteristics of items are bundled in the items. CTR and CVR predicted by the cloud-side models for items are sent to the user's device along with the candidate items. Features related to the user's real-time behavior are produced on the user's device. A summarization of the real-time features is shown in Table~\ref{tab1}. Specifically, the device maintains the candidate items list, once an item is consumed, all features related to it will be collected and fed into mobile models with almost no latency. The real-time features are also uploaded to the server for further analysis and model training. 

\begin{table}[t]
    \centering
    \caption{Features used in edge-side.}
    \begin{tabular}{ccc}
    \hline
    Feature & Source & Description \\
    \hline
    $v_{emb}$     &   & embedding of item ID\\
    $v_{cate}$     &  item   &  item category\\
    $v_{info}$     &  &  statistical information of item,\\
    & & such as monthly turnover\\
    \hline
    $p_{ctr}$ & server & predicted CTR of item\\
    $p_{cvr}$ &  & predicted CVR of item\\
    \hline
    $v_{clk}$ & & clicked item ID\\
    $v_{st}$ & & stay time in an item\\
    $v_{pos}$ & device & item exposure position\\
    $v_f$ & & user's real-time feedback such as click\\
    $v_t$ & & current time: week and hour\\
    \hline
    \end{tabular}
    \label{tab1}
\end{table}

\section{proposed method}
In this section, we describe our Mobile Supply in detail. Going further, we also present our novel mobile ranking considering multi-scene learning. The core idea of the model design is to satisfy the user's interest as accurately as possible in real-time under the limited computing resource.

\begin{figure*}[t]
    \centering
    \includegraphics[width=.82\linewidth]{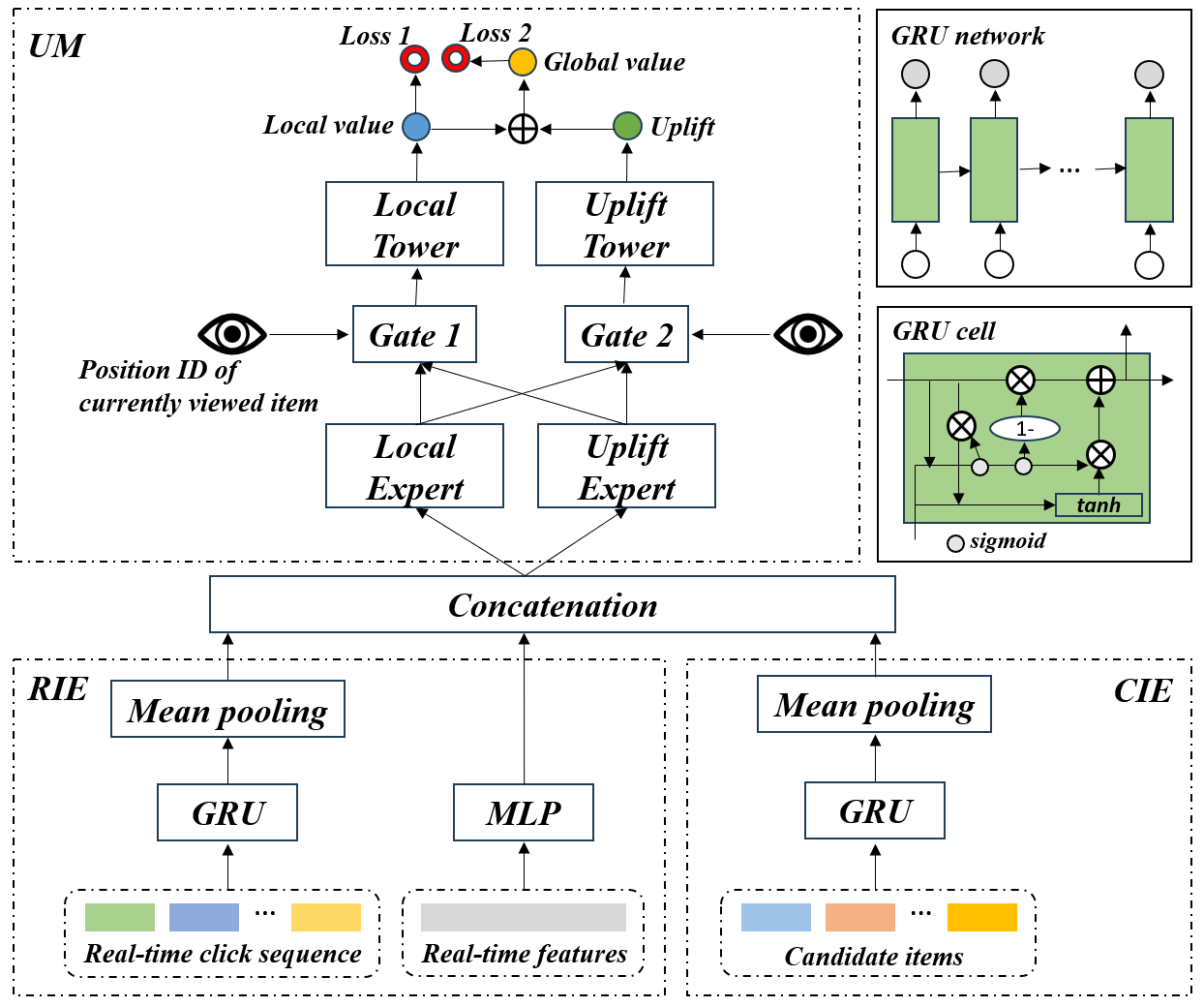}
    \caption{Illustration of Mobile Supply.}
    \label{fig3}
    \Description{The structures of RIE, CIE, UM and GRU.}
\end{figure*}

\subsection{Problem Statement}
Our proposed edge-side recommendation aims to consider the user's real-time behaviors as signals to automatically adjust the recommendation list on the mobile device. Given a ranked item list $L_{sv}$ from the server cached on mobile device and a series of real-time actions $v_f$ of user $u$, our goal is to find two scoring functions: 1) $\Omega_{ms}(\mathbb{I},\mathbb{S},\mathbb{D})$ which can estimate the maximum value uplift that can be achieved by reordering $L_{sv}$. If the uplift is less than the threshold $\alpha$, the server sends a new ranked item list $L_{new}$ to device. 2) $\Omega_{mr}(\mathbb{I},\mathbb{S},\mathbb{D})$ which can compute the value of each item. The items will be re-ranked from largest to smallest according to the value. The two functions consider 
feature sets collected from item, server and device denoted as $\mathbb{I}$, $\mathbb{S}$ and $\mathbb{D}$ respectively. 

\subsection{Mobile Supply}
In this section, we articulate the absolutely new module named Mobile Supply which can automatically send paging requests to further improve the performance of edge-side recommendation. 

\subsubsection{Real-Time Interest Extractor} 
This part is the key point for Mobile Supply to respond for the user's real-time feedback. Here we choose gate recurrent unit (GRU) to encode the real-time click sequence, controlling the update of hidden states with a reset gate and an update gate. Once the real-time click sequence is encoded, we employ mean pooling to obtain the real-time click vector.

All real-time features are fed into an MLP to perform an automatic feature crossing and concatenated with the real-time click vector to generate a real-time information vector. This vector reveals the real-time interest and the real-time status of the user.

\subsubsection{Contextual Information Extractor}
To extract global and local information of the item list, it is necessary to understand the contextual information of the item list. Therefore, we employ GRU+mean pooling to encode the ranked item list to obtain a contextual information vector. This vector reveals the order of the current item list and the effect of the order on the user's decision-making. 

After obtaining the contextual information vector, it is concatenated with the real-time information vector and fed into the uplift MMoE.

\subsubsection{Uplift MMoE}
There are early studies that show that list-wise ranking approach can capture the contextual information well and is more in line with industrial recommendation practices. Traditional point-wise approach considers that the user's behavior is subject to the assumption of independent identical distribution (\textit{i.i.d.}). In fact, the user's behavior is determined by all the items that can be seen on the user's screen. So we need a list-wise paradigm, which takes into account the combination of different items to generate the recommendation list in a global view.

To determine whether Mobile Supply should turn the page, we introduce the concept of value uplift $U_p$ of an item list. The value $\mathcal{V}$ of an item list is defined as the the maximum capacity of an item list that can satisfy the user. The difference between global value and local value is defined as value uplift. 

In a real recommendation system, the number of items in a page is greater than the number of items that can be displayed on the user's device screen. A user is only affected by the items they have seen (exposed items). Here, we define the information in the exposed items as local information of the item list and the value of the list of exposed items as the local value $\mathcal{V}_l$. Similarly, we define the information in items in the current page as global information and the value of list of items in the current page as the global value $\mathcal{V}_g$. The local value describes the degree to which the user is satisfied by the exposed items, while the global value describes the maximum degree to which the items in the current page satisfy the user. If the value uplift is too small, it means that mobile ranking is unlikely to satisfy the user, no matter how it ranks the items in the current page.

Thus, an ideal Mobile Supply can calculate the global value and the local value in a list-wise paradigm. However, it is impossible to deploy a list-wise model in mobile device. The computational complexity of a list-wise model is so high that even deploying it on the server is a challenge. Therefore, we propose an approach to approximate list-wise by a point-wise paradigm. 

Because our ultimate goal is to let users buy items, we use the number of converted items in a list as the label of the value of the list:
\begin{equation}
    \mathcal{V}=\textbf{sum}(\#\textbf{orders}),
\end{equation}
where $\#$ denotes ``the number of".

To estimate the local value and the global value, we propose uplift MMoE considering it as a multi-task learning and the two tasks are obviously relative. All layers in MMoE are DNN. Specifically, the input feature of gates is position ID of currently viewed item $P_{cv}$, which explicitly determines the user's current field of view and allows the model to adapt the weight of information based on this. The output of uplift MMoE can be formulated as the following equations:
\begin{equation}
\begin{split}
    \hat{\mathcal{V}}_l=r(g_1(P_{cv})\cdot [e_l(X), e_u(X)]),\\
    \hat{\mathcal{V}}_g=r(g_2(P_{cv})\cdot [e_l(X), e_u(X)]),
\end{split}
\end{equation}
where $r$ is ReLU, $g_1$ and $g_2$ are functions of gate 1 and 2, $e_l$ and $e_u$ are functions of experts, $X$ is input of UM, and $[]$ means concatenation.

Then, we estimate value uplift as follows:
\begin{equation}
    \hat{U}_p=\hat{\mathcal{V}}_g-\hat{\mathcal{V}}_l.
\end{equation}

In fact, the following formula is always true:
\begin{equation}
\begin{split}
    \mathcal{V}_g\geq\mathcal{V}_l \Rightarrow U_p\geq0.
\end{split}
\end{equation}
Because the list of exposed items is a subset of the list of items in the current page, the local value is a locally feasible solution to the global value.

In order to prevent the model compute a negative uplift in the estimation stage. We consider the uplift as an intermediate variable and trace $\hat{\mathcal{V}}_g$ as follows:
\begin{equation}
    \hat{\mathcal{V}}_g=\hat{U}_p+\hat{\mathcal{V}}_l.
\end{equation}
In this way we can ensure that the estimated uplift is meaningful.

\subsubsection{Model Training}
Because the learning objective is continuous, it is a regression task. Thus the loss function can be formulated as:
\begin{equation}
    \mathcal{L}=\frac{1}{N}\sum_{i=1}^N[(\mathcal{V}_{gi}-\hat{\mathcal{V}}_{gi})^2+(\mathcal{V}_{li}-\hat{\mathcal{V}}_{li})^2],
\end{equation}
where $N$ is the number of samples. This is the general learning objective. However, in our real business, we have observed that there is only one order in most lists. Thus, the special learning objective is:
\begin{equation}
    \mathcal{L}=\mathcal{L}_g+\mathcal{L}_l,
\end{equation}
where $\mathcal{L}_g$ and $\mathcal{L}_l$ are binary cross-entropy loss for $\mathcal{V}_g$ and $\mathcal{V}_l$ and binary cross-entropy loss can be expressed as:
\begin{equation}
    \mathcal{L}(y,\hat{y})=-\frac{1}{N}\sum_{i=1}^N[y_i\log\hat{y}_i+(1-y_i)\log(1-\hat{y}_i)],
\end{equation}
where $y$ is label and $\hat{y}$ is estimator. Especially, in the special learning objective, the value of list indicates whether the current list has at least one order. $\mathcal{V}\in\{0,1\}$ in this case. 

In this paper, we choose the special objective tailored for our business to train Mobile Supply. In this way, we implement a point-wise strategy to consider the impact of list-wise information on users, and achieve a fast and accurate user interest mining.

\subsubsection{Paging Request Sending}
In the inference stage, Mobile Supply calculates the uplift of the current list, and if the uplift is below a threshold $\alpha$, mobile recommender will send a paging request to the server.

\subsection{Device-aware Mobile Ranking}
To match our business scenario and mobile recommendation system with Mobile Supply, we also improve our mobile ranking. Our mobile ranking mainly consists of two parts: 1) real-time interest extractor (RIE) and 2) device MMoE (DM).  

\subsubsection{Proposed Mobile Ranking}
\begin{figure}
    \centering
    \includegraphics[width=0.99\columnwidth]{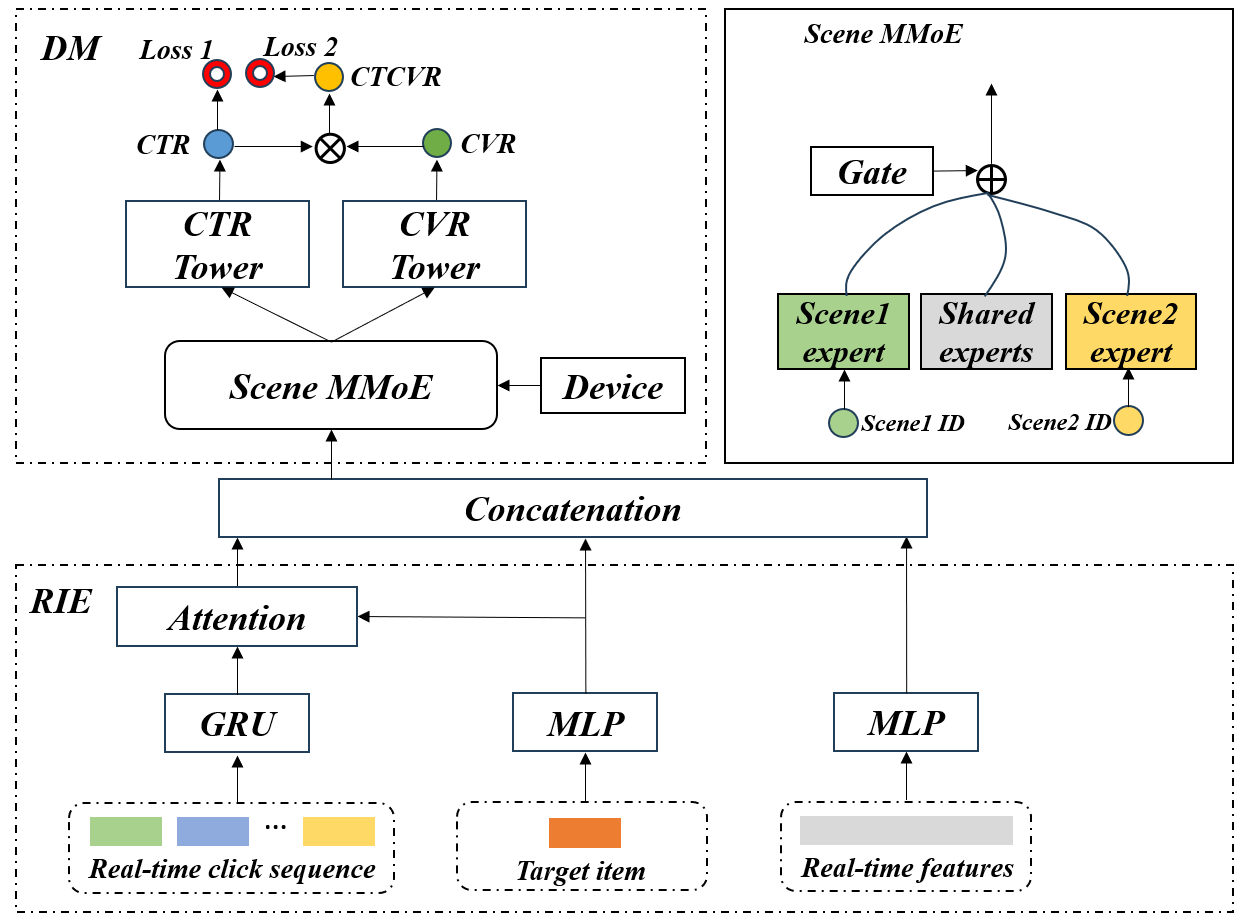}
    \caption{Illustration of Our Proposed Mobile Ranking.}
    \label{fig4}
    \Description{}
\end{figure}
Similar to Mobile Supply, our device-aware mobile ranking is also equipped with a real-time interest extractor which can capture the real-time evolution of user preferences. The real-time click sequence is encoded by a GRU and then a target attention mechanism is used to draw the item vector, expressed as follows.
\begin{equation}
    \text{Attention}(\mathbf{Q},\mathbf{K},\mathbf{V})=\text{softmax}(\frac{\mathbf{Q}\mathbf{K}^T}{\sqrt{d}})\mathbf{V},
\end{equation}
where $\mathbf{Q},\mathbf{K},\mathbf{V}$ are query, key, value, respectively. $d$ is the embedding dimension. The query $\mathbf{Q}$ is projected from features of target item, and $\mathbf{K}$ and $\mathbf{V}$ are both projected from features of click sequence.

Because of the limitation of computing power, we employ MLP to encode the target item and the real-time features. The extracted vectors are concatenated to be input to DM. DM is based on scene MMoE (SMMoE). Here, we consider device type (\textit{i.e.} Android and IOS) as scene ID. SMMoE consists of multiple shared experts and two scene experts. The shared experts are always involved in the inference, but the scene experts will only be activated in the corresponding scenes. For example, the Android expert will only be effective for users using Android phones. Our model can estimate CTR and CVR at the same time. Especially, to avoid sample selection bias issue \cite{ma2018entire}, we use ESMM's idea to estimate CTCVR (Click-Through Conversion Rate):
\begin{equation}
    \mathbb{P}(CTCVR)=\mathbb{P}(CTR)\times\mathbb{P}(CVR).
\end{equation}

Thus the training objective of device-aware mobile ranking is:
\begin{equation}
    \mathcal{L}=\mathcal{L}(CTR)+\mathcal{L}(CTCVR),    
\end{equation}
where $\mathcal{L}(CTR)$ and $\mathcal{L}(CTCVR)$ are binary cross-entropy loss for CTR and CTCVR.

\subsection{Deployment}
We export model checkpoints periodically in the training process, then convert the checkpoint to TFLite format, and convert it to MNN file in Jarvis environment. If the edge-side model is outdated, then the server tells the device to upgrade APP and download the new models.

\section{experiments}
We conduct extensive experiments to evaluate the performance of our proposed Mobile Supply and the following research questions (RQs) are answered:

\begin{itemize}
    \item \textbf{RQ1} Does "Mobile Supply+mobile ranking" better than single mobile ranking in mobile recommender systems?
    \item \textbf{RQ2} How do important parts affect the performance of Mobile Supply?
    \item \textbf{RQ3} How does critical parameter $\alpha$ affect the performance of Mobile Supply?
    \item \textbf{RQ4} Does Mobile Supply work in real large-scale online recommendation scenarios?
\end{itemize} 

\subsection{Experimental Settings}
\subsubsection{Datasets}\footnote{To the best of our knowledge, there are no public datasets suited for this task.}
We collect two offline datasets by collecting the Android-phone and iPhone (IOS) users’ feedback logs from a large-scale online platform serving billions of users. The statistics of offline datasets are listed in Table~\ref{tab2}.

\begin{table}[t]
    \centering
    \caption{Description of datasets. \# means ``the number of".}
    \begin{tabular}{cccccc}
    \hline
    & \#users & \#items & \#clicks & \#purchases & total size \\
    \hline
    Android &  1.98M & 2.16M & 10.52M & 1.00M & 186.56M \\
    IOS     & 1.97M & 2.52M & 11.44M & 1.16M & 237.12M   \\
    \hline
    \end{tabular}
    \label{tab2}
\end{table}

\subsubsection{Metrics and Mobile Ranking Baselines}
We use AUC of each target as evaluation metric. The representative state-of-the-art (SOTA) mobile ranking baselines are listed as follows and the descriptions of them are shown in Table~\ref{tab3}.
\begin{itemize}
    \item \textbf{EdgeRec} \cite{gong2020edgerec} is deployed on Taobao, an e-commerce application. We implement it as described in the paper with real-time features and feature engineering. 
    \item \textbf{KuaiRec} \cite{gong2022real} is deployed on Kuaishou, a short-video application. We implement it as described in the paper with real-time features and feature engineering.
    \item \textbf{DMR} is deployed on an online food platform serving billions of users. This is our proposed mobile ranking technique in this paper.
\end{itemize}

\begin{table}[t]
    \centering
    \caption{Descriptions of mobile rankings. FE is feature encoder, MS is main structure, \# means ``the number of", and IOP means ``in one page".}
    \begin{tabular}{ccccc}
    \hline
     & FE & MS & \#parameters & \#items IOP\\
    \hline
     EdgeRec    & GRU & MLP & 20K &50 \\ 
     KuaiRec    & MHA & MMoE & 140K & 10 \\
     DMR        & GRU/MLP & SMMoE & 37K & 20 \\
    \hline
    \end{tabular}
    \label{tab3}
\end{table}

\subsubsection{Training Protocol}
Models in this paper are implemented with Tensorflow 1.12 in Python 2.7 environment. The models are trained with a chief-worker distributed framework with 1600 CPUs. The initial learning rate is set to 0.005 and the optimizer is set to AdagradDecay. Besides, batch size is set to 1024 and training epoch is set to 1 because of one-epoch phenomenon \cite{zhang2022towards}.

\subsection{RQ1: Controlled Trials}
Here, we conduct controlled trials to evaluate the functionality of Mobile Supply. There are three pairs of comparison experiments, depending on whether Mobile Supply is activated or not. Table~\ref{tab4} and~\ref{tab5} indicate that Mobile Supply can further enhance all SOTA mobile ranking baselines in terms of CTR-AUC and CTCVR-AUC on two real-world datasets. Besides, the improvements prove that our proposed approaches achieve a significant improvement\footnote{Note that the 0.1\% AUC gain is already considerable in large-scale industrial recommender \cite{du2022basm, jiang2023esmc}.} for mobile recommender systems. Thus, Mobile Supply is a meaningful and promising module in the pipeline of recommenders.

For mobile ranking, our proposed DMR achieves the best performance. This is mainly because we take into account the differences between people who use different devices. User behavior logs have a large gap in data distribution between Android and IOS phones, which makes it challenging to mix them together for modeling. Through a multi-scene learning strategy, we can decouple these two scenes to a certain extent, learn specific model parameters, and improve the performance.

In terms of dataset, Mobile Supply brings much more improvement on IOS than it does on Android, and the difference is about five times! This further illustrates the differences in the behavior of Android and IOS users, and demonstrates the necessity of using DMR as mobile ranking.

\begin{table}[t]
    \centering
    \caption{Results of controlled trials on Android dataset. Improvement is calculated as the relative increase of the performance of mobile ranking+Mobile Supply compared to that of single mobile ranking.}
    \begin{tabular}{ccc}
    \hline
     & CTR-AUC & CTCVR-AUC  \\
    \hline
     EdgeRec    & 0.70006 & 0.81732 \\
     EdgeRec+Mobile Supply & 0.70139 & 0.81839 \\
     Improvement & +0.19\% & +0.13\% \\
     \hline
     KuaiRec    & 0.69807 & 0.81746 \\
     KuaiRec+Mobile Supply & 0.69886 & 0.82363 \\
     Improvement & +0.11\% & +0.75\% \\
     \hline
     DMR   &  0.73125 & 0.82931 \\
     DMR+Mobile Supply & 0.73211 & 0.83076\\
     Improvement & +0.12\% & +0.17\% \\
     \hline
    \end{tabular}
    \label{tab4}
\end{table}

\begin{table}[t]
    \centering
    \caption{Results of controlled trials on IOS dataset. Improvement is calculated as the relative increase of the performance of mobile ranking+Mobile Supply compared to that of single mobile ranking.}
    \begin{tabular}{ccc}
    \hline
     & CTR-AUC & CTCVR-AUC  \\
    \hline
     EdgeRec    & 0.72183 & 0.81105 \\
     EdgeRec+Mobile Supply & 0.72643 & 0.81748 \\
     Improvement & +0.64\% & +0.79\% \\
     \hline
     KuaiRec    & 0.72051 & 0.80939 \\
     KuaiRec+Mobile Supply & 0.72560 & 0.81803 \\
     Improvement & +0.71\% & +1.07\% \\
     \hline
     DMR    & 0.75142 & 0.82407 \\
     DMR+Mobile Supply & 0.75553 & 0.82817 \\
     Improvement & +0.55\% & +0.50\% \\
     \hline
    \end{tabular}
    \label{tab5}
\end{table}

\begin{figure}[htbp]
	\centering
	\subfloat[CTR-AUC on Android.]{\includegraphics[width=.45\columnwidth]{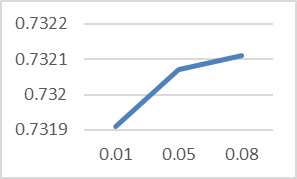}}\hspace{5pt}
	\subfloat[CTCVR-AUC on Android.]{\includegraphics[width=.45\columnwidth]{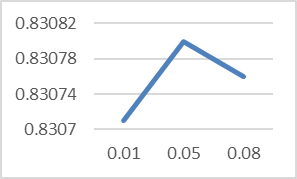}}\\
	\subfloat[CTR-AUC on IOS.]{\includegraphics[width=.45\columnwidth]{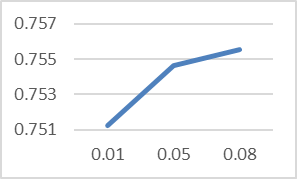}}\hspace{5pt}
	\subfloat[CTCVR-AUC on IOS.]{\includegraphics[width=.45\columnwidth]{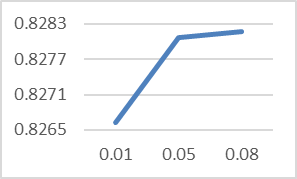}}
	\caption{Sensitivity test of threshold $\alpha$.}
 \label{fig5}
\end{figure}

\subsection{RQ2: Ablation Study}
Here, we study the effects of RIE and CIE on Mobile Supply. Table~\ref{tab6} shows the results of ablation study. The results show that both RIE and CIE can improve the performance of Mobile Supply significantly. We observe that when RIE is removed, uplift AUC would fall by 13\% on average! RIE introduces real-time behaviours and features to models which are critical for real-time recommendation. When CIE is removed, uplift AUC would fall by 3.3\% on average! CIE introduces contextual information to models which is important to capture the browsing status of user. Evidently, it is necessary to employ RIE and CIE in Mobile Supply. In addition, RIE is more important for Mobile Supply and the global value is more easily influenced by the model structure.

\begin{table}[t]
    \centering
    \caption{Average AUC of ablation study on two datasets, ``w/o" is ``without".}
    \begin{tabular}{cccc}
    \hline
    & local value & global value & uplift \\
    \hline
    Mobile Supply w/o RIE & 0.76758 & 0.68151 & 0.67534 \\
    Mobile Supply w/o CIE & 0.80889 & 0.78058 & 0.77241  \\
    Mobile Supply & 0.82597 & 0.81345 & 0.80529 \\
    \hline
    \end{tabular}
    \label{tab6}
\end{table}

\subsection{RQ3: Sensitivity of Parameter Test}
Fig.~\ref{fig5} illustrates the curves of model performance varying with the threshold $\alpha$. The curves show a trend of first growth, then slowing down and then declining. Especially, Mobile Supply generally performs well when $\alpha$ being set to 0.05 in our experiments. A large threshold will cause Mobile Supply to send paging requests frequently, and some valuable items will be missed by the user. A small threshold would disable Mobile Supply. As a result, we need to adjust this parameter carefully when using Mobile Supply. Fortunately, Mobile Supply can bring significant returns in most cases.

\subsection{RQ4: Online A/B Test}
\begin{table}[t]
    \centering
    \caption{Results of five-day online A/B test.}
    \begin{tabular}{ccccccc}
    \hline
         & Day1 & Day2 & Day3 & Day4 & Day5 & Avg. \\
    \hline
     NU   & +0.78\% & +0.82\% & +0.26\% & +0.47\% & +0.75\% & +0.62\%  \\
     NO   &  +0.53\% & +1.07\% & +0.13\% & +0.45\% & +0.86\%  & +0.61\% \\
     OR   &  +0.53\% & +0.95\% & -0.03\% & +0.31\% & +0.75\% & +0.50\% \\
     NG   &  +0.46\% & +0.66\% & -0.61\% & +0.15\% & +0.59\% & +0.26\% \\
    \hline
    \end{tabular}
    \label{tab7}
\end{table}
From August 01, 2023 to August 05, 2023, we conducted a five-day online experiment by deploying DMR+Mobile Supply to the recommendation scenario on the homepage of our online platform. The online base model is a single DMR without Mobile Supply. Here, we select four business-related metrics: Number of Paying Users (NU), Number of Orders (NO), Order Rate (OR) and Net GMV (NG, a measure of net profit) to evaluate the performance in online environment. Results of strictly online A/B test are shown in Table~\ref{tab7}. We can see that the proposed Mobile Supply significantly improves the mobile recommender. On average, our method improves NU by 0.62\%, NO by 0.61\%, OR by 0.50\%, and NG by 0.26\% compared to the base model. This demonstrates the effectiveness of Mobile Supply in large-scale online recommendation system. Our proposed approaches have been deployed on the homepage of our large-scale online food platform serving more than one billion recommendation requests per day.

\begin{figure}[t]
    \centering
    \includegraphics[width=0.98\columnwidth]{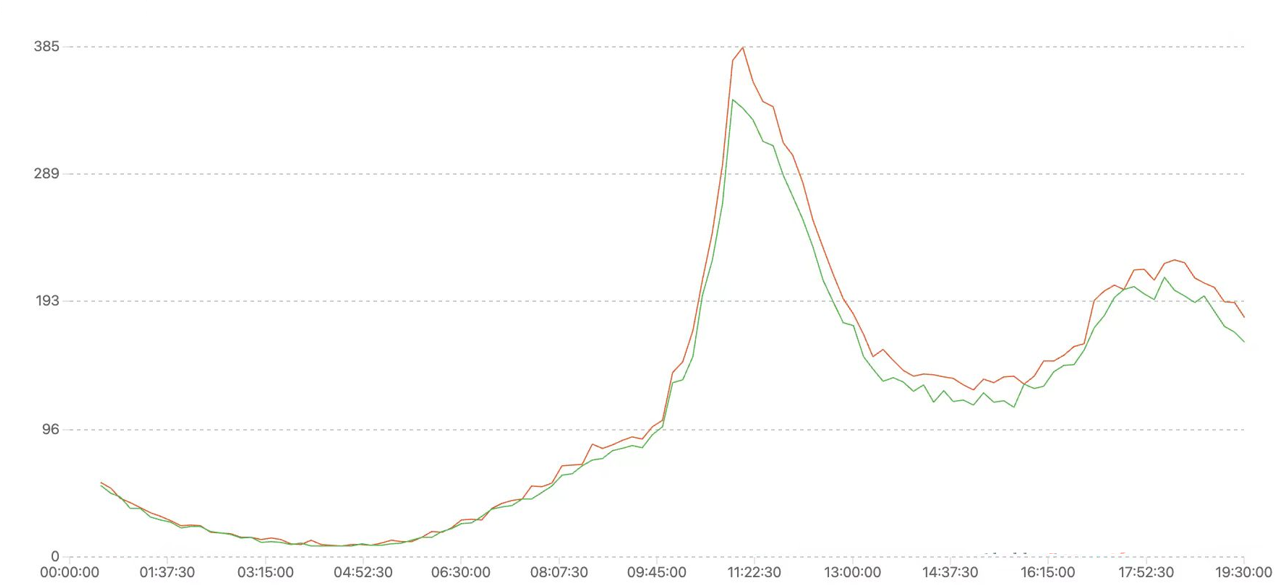}
    \caption{Curve of server stress testing.}
    \label{fig6}
\end{figure}

Without Mobile Supply, users need to send paging requests manually while Mobile Supply can send paging requests automatically based on user behaviors. If Mobile Supply can accurately capture user intent, its frequency of sending paging requests should be close to that of user. Fig.~\ref{fig6} illustrates the curve of server stress testing. The vertical axis indicates QPS of the server, a unit for server stress that is in direct proportion to the frequency of receiving paging requests from the edge-side. The case without Mobile Supply is in green and the case employing Mobile Supply is in red. As we can see, the two curves are very similar in different time periods. This proves that Mobile Supply can accurately capture users' paging intentions, further proving its effectiveness.

\section{conclusion}
In this paper, we propose a novel module in the pipeline of recommender systems named Mobile Supply, as well as a new mobile ranking approach for edge-side recommendation scenarios, to make the full use of the real-time user interest and provide the user an immersive experience. The pipeline of recommender system becomes ``retrival->pre-ranking->ranking->re-ranking->Mobile Supply-> mobile ranking". We specifically design two small model architectures that can be directly and completely deployed on mobile devices, and can perform well under the limitations of computation power. We introduce a point-wise approximation of list-wise to effectively capture user interest. The whole framework is tested both offline and online in a billion-user scale takeaway application, and the results verify its superiority. In the future, we will explore how to upgrade the collaboration between mobile ranking and Mobile Supply to further improve user experience.


\bibliographystyle{ACM-Reference-Format}
\bibliography{ref}


\end{document}